\documentclass[aps,pre,twocolumn,superscriptaddress,showpacs,showkeys,floatfix,a4paper]{revtex4-1}
\usepackage{amsmath}
\usepackage{amsfonts}
\usepackage{graphicx} 
\usepackage{amssymb} 
\usepackage{grffile}
\usepackage{float}
\usepackage{color}
\usepackage[T1]{fontenc}

\newcommand{\SI}{Appendix}

\begin{document}
\title{Skewness of local logarithmic exports}

\author{Sung-Gook Choi}
\affiliation{Department of Physics, Inha University, Incheon 22212, Korea}
\affiliation{Korea Housing \& Urban Guarantee Corporation, Busan 48400, Korea}
\author{Deok-Sun Lee}
\affiliation{School of Computational Sciences, Korea Institute for Advanced Study, Seoul 02455, Korea}
\email{deoksunlee@kias.re.kr}

\date{\today}

\begin{abstract}
The distributions of trade values and relationships among countries and product categories reflect how countries select their trade partners and design export portfolios. Here we consider the exporter-importer network and the exporter-product network with directed links weighted by the logarithm of the corresponding export values each year from 1962 to 2018, and study how the weights of the outgoing links from each country are distributed. Such local logarithmic export distributions by destinations and products are found to follow  approximately the Gaussian distribution across exporters and time, implying random assignment of export values on logarithmic scale. However, a non-zero skewness is identified, changing from positive to negative as exporters have more partner importers and more product categories in their portfolios.  Seeking the origin, we analyze how local exports depend on the out-degree of exporter and the in-degrees of destinations/products and formulate  their quantitative and measurable relation incorporating randomness, which uncovers the fundamental nature of the export strategies of individual countries.
\end{abstract}

\maketitle

\section{Introduction}
\label{sec:intro}

Individual countries, though equipped with different resources for production and put in different environments, commonly  fine tune the strategies on how to allocate resources, invest capital, and make international trade partnership towards the economic development~\cite{Hidalgo482,Tacchella:2018aa,10.2307/1926806,kuznets1971economic,10.1371/journal.pone.0129955,Choi:2019aa}.  Such efforts are universal and thus can be imprinted in the patterns of domestic production, consumption, and international trade.  Especially in international trade are reflected the decisions on what products, to what countries, and how much each country exports~\cite{10.1257/0002828054201396}. The empirical and analytical study of the trade value distributions may therefore illuminate  the underlying principles of the global economic systems. 

The network approach has been instrumental to quantify various nature of international trade, revealing its far-from-random connection among countries~\cite{PhysRevE.68.015101,PhysRevLett.93.188701,PhysRevE.79.036115}, and  the broadness of the distributions of trade values by countries~\cite{cha2010,1367-2630-17-1-013009,PhysRevE.95.052319} or products~\cite{hidalgo2009building,tacchella2012new,1742-5468-2017-12-123403}. These studies mostly aimed at identifying typical representative characteristics, consequently neglecting local variations which may be essential for understanding the principles governing the trade strategies of individual countries. In this light, here we investigate how the export values of a country are distributed by its destinations (partner importers) and product categories. We compile the international trade data-sets from two sources~\cite{NBERw11040,wto2020}, covering different time periods, to obtain the annual export value for each pair of an exporter and an importer country, and for each pair of an exporter country and a product category. Using the result, we construct the exporter-importer (EI) network and the exporter-product (EP) network each year, where links are directed, from exporter countries to their partner importers or to their exported product categories, and are weighted by the logarithm of the corresponding annual export values. Then the distribution of the weights of the outgoing links from each country represents how the values of the exports from the country are distributed by destinations or products, which we analyze in the present study. 

We find that such local logarithmic export distributions measured for individual exporters follow approximately the Gaussian distribution, implying the possibility of the first two moments  to fully characterize the whole distributions.  However, the distributions exhibit significant asymmetry and significant non-zero skewness. Moreover, the skewness shows a systematic variation - decreases with the out-degree of exporter, the number of partner importers or of the exported products. In seeking the origin, we first find that the average export of a country per destination/product increases with its out-degree, meaning that the whole distribution is pushed relatively to the right (left) for a hub (non-hub) exporter and thus may give rise to a thicker left (right) tail.  From the network perspective, this finding characterizes the influence of a source node on the average weight of its outgoing links.  Individual link weights are found to be dependent also on the in-degrees of the target nodes,  the number of connected exporters of destination/product  in the EI/EP network. Yet the dependence is accompanied by large fluctuations. We derive the formula decomposing a local export into a deterministic term dependent on the in-degree of the target and a random noise, which  is well defined empirically and leads us to disentangle the skewness of local exports in terms of the third order joint cumulants of the in-degrees of the targets and random noises.  These results therefore uncover the mechanisms generating the skewed distribution of local exports but also  provide a quantitative framework to analyze the architecture of weighted complex networks. 

The paper is organized as follows. In Sec.~\ref{sec:empirical}, we introduce briefly the data-sets and present the empirical findings on  the skewness of the local logarithmic export distributions.  We study in Sec.~\ref{sec:indegree} the distributions of the in-degrees of destinations and products receiving links from each exporter and compare the skewness of such local in-degrees and local exports. In Sec.~\ref{sec:decomposing}, we show how the in-degrees of destinations/products and  random noises together generate the skewness of local exports. We discuss our findings and  future works in Sec.~\ref{sec:discussion}.

\section{Local export distributions}
\label{sec:empirical}

\subsection{Data-sets and two trade networks}
We compile the NBER-UN data~\cite{NBERw11040} and the World Trade Organization (WTO) data~\cite{wto2020} to obtain the annual export value $V_{cpc^\prime}(t)$ in US dollars of a product category  $p$ exported by a country $c$ to another country $c'$ in the period $1962\leq t\leq  2018$. The number of  countries exporting or importing at least one product ranges between 154 and 190 and the total number of exported products  fluctuates between 686 and 1458 in the studied period  [\SI~\ref{sec:datasets}]. 

Each country exports selected products to selected destination countries in different amounts, which constitutes its export portfolio.  To represent the whole collection of such trade relationships, here we consider two kinds of directed and weighted  networks, the exporter-importer(EI) network and the exporter-product(EP) network each year as follows [Fig.~\ref{fig:network}]. Nodes  in the EI network are countries exporting or importing at least one product in given year $t$.  A link is assigned from a country $c$ to another country $c^\prime$ if the aggregated export from $c$ to $c^\prime$, $V_{cc^\prime}(t) = \sum_p V_{cpc^\prime}(t)$, is positive. The (binary) adjacency matrix  ${\bf A}^{\rm (EI)} = (A_{cc^\prime})$ is given by $A_{cc^\prime} = \theta(V_{cc^\prime})$ with $\theta(x)=1$ for $x>0$ and $0$ for $x\leq 0$. The EP network is a bipartite network consisting of two types of nodes, countries and product categories, and they are connected if $V_{cp}(t) = \sum_{c^\prime} V_{cpc^\prime}(t)>0$. Therefore the adjacency matrix ${\bf A}^{\rm (EP)}= (A_{cp})$ is given by  $A_{cp} = \theta(V_{cp})$. The links in the EI and the EP network are then weighted by the base-10 logarithm of the corresponding export values as
\begin{align}
v_{cc^\prime}(t) &= \log_{10}\left(V_{cc^\prime}(t)\right), \nonumber\\
v_{cp}(t) &=\log_{10}\left(V_{cp}(t)\right).
\label{eq:v}
\end{align}
The link-weight matrix, or the weighted adjacency matrix, is given by ${\bf v}^{\rm (EI)} = (v_{cc^\prime})$ and ${\bf v}^{\rm (EP)} = (v_{cp})$, respectively.   We take logarithm given that export values span several orders of magnitude~\cite{1742-5468-2008-02-P02002,PhysRevE.79.036115,cha2010,PhysRevE.95.052319}. As we will show, the distributions of $v$'s are close to the Gaussian distribution and we will investigate their characteristics. On the other hand, the distribution of the raw values of export $V$'s are close to power-laws~\cite{cha2010,PhysRevE.95.052319}. 

\subsection{Distribution of local logarithmic exports}

How a country $c$ divides its export among different destinations ($c^\prime$) or product categories ($p$) can be represented by the distributions of its logarithmic export values 
\begin{align}
P_c^{\rm (EI)} (v) &=  {1\over q_{c}^{\rm (EI)}} \sum_{c^\prime \in \mathtt{I}_c^{\rm (EI)}} \delta(v_{cc'} - v),\nonumber\\
P_c^{\rm (EP)} (v) &= {1\over q_{c}^{\rm (EP)}} \sum_{p\in \mathtt{I}_c^{\rm (EP)}} \delta(v_{cp} - v),
\label{eq:Pcvdef}
\end{align}
where $\mathtt{I}_c^{\rm (EI)}=\{c^\prime | A_{cc^\prime}>0\}$ and $\mathtt{I}_c^{\rm (EP)} = \{ p|A_{cp}>0\}$ are the set of the partner importers and of the exported products of $c$. The out-degree $q_c^{\rm (EI)}$ and $q_c^{\rm (EP)}$ are the size of those sets equal to the number of outgoing links from $c$,
\begin{align}
q_c^{\rm (EI)} &=\sum_{c^\prime} A_{cc^\prime}= \sum_{c^\prime \in \mathtt{I}_c^{\rm (EI)}} 1,\nonumber\\
q_c^{\rm (EP)} &=\sum_p A_{cp}= \sum_{p \in \mathtt{I}_c^{\rm (EP)}} 1.
\end{align}
 $q_c^{\rm (EI)}$ and $q_c^{\rm (EP)}$ can be considered as a measure of the diversity of country $c$ in the trade partnership and inventory. $P_c^{\rm (EI)} (v)$ and $P_c^{\rm (EP)} (v)$ correspond to ego-centric link-weight distributions in the EI and the EP network and we call  them {\it local} (logarithmic) export distributions in that they are measured only for the outgoing links of $c$.

\begin{figure}
\includegraphics[width=\columnwidth]{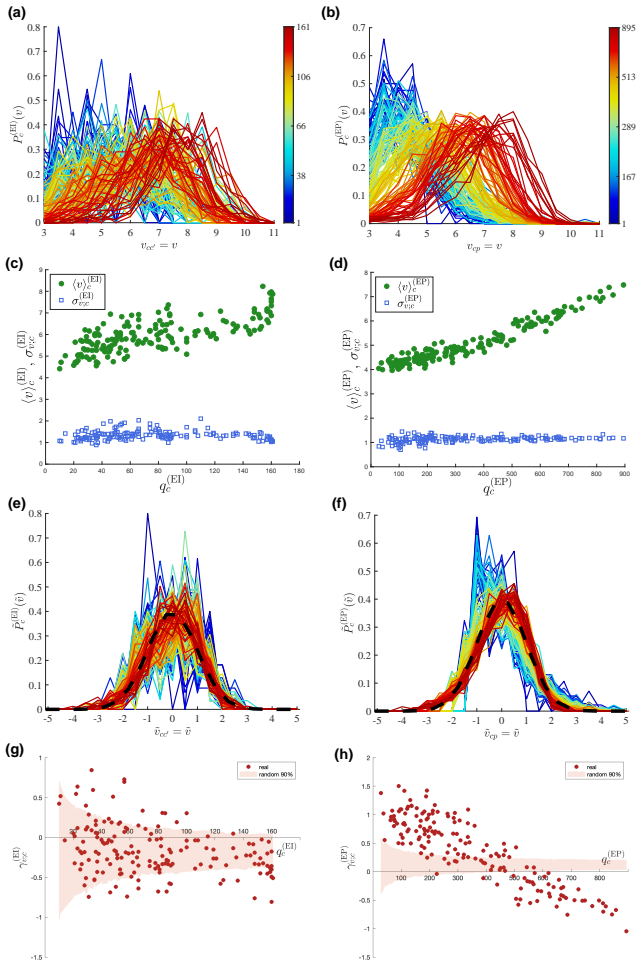} 
\caption{
{\bf Statistics of the logarithm of local exports.}
(a, b) Distribution of the logarithm of the exports from each country $c$ (a) to different destinations and (b) of different product categories  in year 1980. Color varies with the out-degree of $c$. 
(c, d) The mean  and standard deviation  of the  logarithmic exports from country $c$ (c) to different destinations and (d) of different products  versus the out-degree of $c$.  
(e, f) Distribution of the standardized logarithmic exports from country $c$ (e) to different destinations  and (f) of different products. 
The dashed lines  in (e) and (f) are $\tilde{P}^{\rm (Gaussian)}(\tilde{v}) = {e^{-{\tilde{v}^2\over 2}}\over \sqrt{2\pi}}$.  
(g, h) The skewness of the logarithmic exports from $c$ (g) to different destinations and (h) of different products versus the out-degree of $c$. Shades in (g) and (f) represent the middle 90\% range of the random-version skewness.
} 
\label{fig:Pcv}
\end{figure}

The local export distributions $P_c(v)$ commonly  take a bell shape while the mean  $\langle v\rangle_c^{\rm (EI)}$ or $\langle v\rangle_c^{\rm (EP)}$ is different from exporter to exporter [Fig.~\ref{fig:Pcv} (a-d)]. We present the results obtained for year 1980 as an example in most figure plots,  but all the statements of ours are valid throughout the studied period unless stated otherwise. The standard deviation $\sigma_{v;c}^{\rm (EI)}$ or $\sigma_{v;c}^{\rm (EP)}$ does not vary with $c$ as significantly as the mean does.  Let us introduce the standardized variable $\tilde{v}$ such that 
\begin{equation}
v_{c\nu} = \langle v\rangle_c^{\rm (EG)} + \sigma_{v;c}^{\rm (EG)} \tilde{v}_{c\nu} 
\ {\rm with} \ (\nu,{\rm G})
= \left\{\begin{array}{l}
 (c^\prime, {\rm I}) \ {\rm or} \\
 (p, {\rm P})
 \end{array}\right.
\label{eq:vtilde}
\end{equation}
where  we used $\nu$ and $G$ to avoid repeating almost the same expressions such that $\nu$ represents an importer $c^\prime$ or a product $p$, and G represents I or  P.  Then the standardized distributions $\tilde{P}_c(\tilde{v})$ for different  $c$'s look collapsing onto the Gaussian function of zero mean and unit variance $\tilde{P}^{\rm (Gaussian)}(\tilde{v}) = {e^{-{\tilde{v}^2\over 2}}\over \sqrt{2\pi}}$ [Fig.~\ref{fig:Pcv} (e) and (f)].  If the data collapse were perfect, $\{\tilde{v}_{c\nu}|\nu\in \mathtt{I}_c\}$ would be considered as Gaussian random variables satisfying  $\langle \tilde{v}\rangle_c = 0, \ \langle \tilde{v}^2\rangle_c = 1, \ \langle \tilde{v}^n\rangle_c =0$ for  $n\geq 3$ with $\langle \tilde{v}^n \rangle_c ={1\over q_c} \sum_{\nu\in\mathtt{I}_c} \tilde{v}^n_{c\nu}$. We would think that every country divides its export among destinations and products randomly on logarithmic scale universally and that the mean and the standard deviation fully characterize the statistics of the  local logarithmic export values.

\subsection{Skewness}

The data collapse is not perfect, however. To quantify the deviation, we measure the skewness of local logarithmic exports for each exporter $c$ as  
\begin{equation}
\gamma_{v;c}  = 
{{1\over q_c} \sum_{\nu\in \mathtt{I}_c}  (v_{c\nu} - \langle v\rangle_c)^3  \over {\sigma_{v;c}}^3} 
=\langle \tilde{v}^3\rangle_c,
\end{equation}
where we dropped the superscript (EI) and (EP) but they are assumed implicitly. The results are in Fig.~\ref{fig:Pcv} (g) and (h). Skewness represents the normalized asymmetry of a distribution such that the skewness is positive (negative) if the right(left) tail is thicker.   The observed $\gamma_{v;c}^{\rm (EP)}$'s are significant for most exporters $c$, and $\gamma_{v;c}^{\rm (EI)}$'s seem partly so. To check the significance, we shuffle weights - logarithmic export values - among links to obtain a thousand of realizations of weight-shuffled EI and EP network and measure their skewness.   In case of year 1980 presented in Fig.~\ref{fig:Pcv} (g) and (h),  140 among 161 $(=87\%)$ exporters with out-degrees not smaller than 10  find their true skewness $\gamma_{v;c}^{\rm (EP)}$ outside the middle 90\% range of the skewness from randomized networks  [Fig.~\ref{fig:Pcv} (h)].  42 ($=26\%$)  exporters have true skewness $\gamma_{v;c}^{\rm (EI)}$ outside the middle 90\% range of the random-version skewness [Fig.~\ref{fig:Pcv} (g)].  See Appendix~\ref{sec:sigcorr} and Fig.~\ref{fig:sig} for more details.

The skewness of local logarithmic exports tends to decrease, from positive to negative, with the out-degree of the exporter $c$, more strongly for $\gamma_{v;c}^{\rm (EP)}$ than for $\gamma_{v;c}^{\rm (EI)}$ [Fig.~\ref{fig:Pcv} (g) and (h)]. Such correlation is robustly observed in the considered period, except for the weak correlation between $\gamma_{v;c}^{\rm (EI)}$ and $q_c^{\rm (EI)}$  in the middle eighties and nineties [Fig.~\ref{fig:corrskewdiversity}].  

When  the distribution of a finite-ranged variable is much closer to  its lower (upper) bound,  a thicker right (left) tail can be developed, yielding  a positive (negative) skewness. This can explain the observed negative correlation between $\gamma_{v;c}$ and $q_c$; $P_c(v)$ moves to the right on the $v$ axis if $c$ has a large out-degree, as captured by the growth of the mean logarithmic export $\langle v\rangle_c$ with $q_c$ [Fig.~\ref{fig:Pcv} (a-d)].  Empirically $\gamma_{v;c} = \langle \tilde{v}^3\rangle_c$ is negatively correlated with $\langle v\rangle_c$ as expected  [Fig.~\ref{fig:skewness_mvc}]. It is noteworthy that the export value of a country per destination/product grows with the out-degree of the exporter, as it implies  a super-linear growth of the whole export of a country $V_c = \sum_{c^\prime, p} V_{cc^\prime p}$ with  the out-degree $q_c^{\rm (EI)}$ and $q_c^{\rm (EP)}$. More developed or richer countries may have many  partners and export products~\cite{1742-5468-2008-02-P02002,cha2010,1367-2630-17-1-013009,PhysRevLett.93.188701} and therefore our results point out different skewness of export portfolio between developed and developing countries. 

We have just argued that the skewness of local exports $\gamma_{v;c}$  are influenced by the exporter $c$ via the $c$-dependent mean $\langle v\rangle_c$.  Individual local exports $v_{cc^\prime}$'s and $v_{cp}$'s must be dependent also on  the importer country $c^\prime$ and the product category $p$~\cite{10.1371/journal.pone.0129955}. Exporters have different sets of partners and export products, possibly depending on the exporters' out-degrees,  which might be influencing the skewness of local export distributions. In the next sections, we study how the standardized logarithmic exports $\tilde{v}_{c\nu}$ vary with $\nu$, which will deepen our understanding of the statistics of local exports and of the export strategies of countries in international trade.

\section{Local in-degree distributions}
\label{sec:indegree}

What products and to what countries a country exports and how large its value is may depend on  the domestic industrial landscape of the exporter itself but also on the characteristics of the partner importers and product categories such as the economic development of and the social and political relationship with the partner countries, and the technology level of products, and so on. Among them, how many countries export to a country and how many countries export a product category can be used as important information when an exporter country constructs its product portfolio, as they represent the {\it popularity} in international trade, reflecting the collective evaluation by countries~\cite{10.1371/journal.pone.0129955}. They correspond to the in-degrees of a country $c'$ and of product $p$ in the EI and the EP network, defined as  
\begin{equation}
k_{\nu}  = \sum_c A_{c\nu}=\sum_{c\in \mathtt{O}_{\nu}} 1, 
\end{equation}
with $\nu$ being $c^\prime$ or $p$, and  $\mathtt{O}_{c^\prime/p}$ the set of nodes (countries) sending outgoing links to $c^\prime$ or $p$. We take the in-degree as the representative characteristic of an importer or a product and investigate how the local logarithmic export value to/of a destination/product relates to the in-degree of the latter.  To this end, we examine in this section the distributions of the in-degrees of the partner importers and the exported products of given exporter, which we call local in-degree distributions,  and their correlations, if any, with local export values. 

\begin{figure}
\includegraphics[width=\columnwidth]{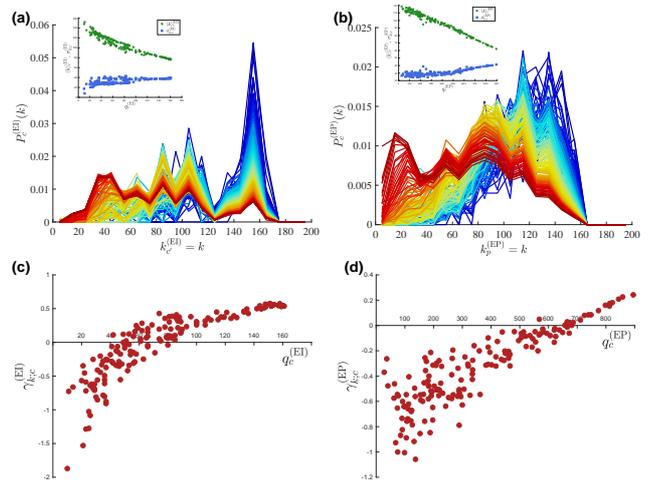}
\caption{{\bf Statistics of local in-degrees in the EI and the EP network.} The in-degree of a product or a country node  means the number of exporter countries exporting the product or to the country.
(a, b) Distribution of the in-degrees of (a) the country nodes in the EI network and (b) the product nodes in the EP network receiving links from each exporter country $c$  in year 1980.  Color varies with the out-degree  of $c$. Inset: The mean  and standard deviation  of local in-degrees versus the out-degree of exporter.
(c, d) The skewness of the in-degrees of (c) the country in the EI network and (d) the product nodes in the EP network receiving links from $c$ versus the out-degree of the exporter $c$.} 
\label{fig:indegreedist}
\end{figure}

\subsection{Local in-degree distribution and its moments}

Like the local export distributions, the local in-degree distributions  $P_c^{\rm (EI)}(k) = {1\over q_c^{\rm (EI)}}\sum_{c^\prime \in \mathtt{I}_c^{\rm (EI)}} \delta(k_{c^\prime}^{\rm (EI)} - k)$ and $P_c^{\rm (EP)}(k) = {1\over q_c^{\rm (EP)}}\sum_{p \in \mathtt{I}_c^{\rm (EP)}} \delta(k_{p}^{\rm (EP)} - k)$ show a large variation with exporters $c$,  related to the variation of the mean local in-degree $\langle k\rangle_c$ [Fig.~\ref{fig:indegreedist} (a) and (b)].  Differently from the mean local logarithmic export,  $\langle k\rangle_c$ decreases with the out-degree  of exporter. The standard deviations $\sigma_{k;c}^{\rm (EI)}$ and $\sigma_{k;c}^{\rm (EP)}$ increase with $q_c$.  Such behaviors of the first two moments and the whole distributions  imply that hub exporters select unpopular destinations/products as well as popular ones while non-hub exporters tend to select popular partners and products only. In the context of complex networks, hub exporters and popular destinations/products have links to and from one another, forming a core, while non-hub exporters and unpopular destinations/products  tend to have links only to and from the nodes in the core, viewed as periphery, which is characteristic of the core-periphery structure~\cite{PhysRevE.89.032810}.

The skewness of the local in-degree distribution $\gamma_{k;c}  = {1\over q_c} \sum_{\nu\in \mathtt{I}_c} (k_{\nu} - \langle k\rangle_c)^3/\sigma_{k;c}^3$ increases with the out-degree of exporter as shown in Fig.~\ref{fig:indegreedist} (c) and (d).  See Fig.~\ref{fig:corrskewindeg} as well for its robustness against time.  As we discussed for the correlation of the skewness of local export and the out-degree of exporter,  the growth of $\gamma_{k;c}$, with $q_c$ may be caused by the shift of local in-degree distributions in the $k$ axis to the left as the out-degree of exporter increases. Also the distributions are narrower for non-hub exporters, explaining the larger absolute values of  $\gamma_{k;c}$ for non-hub exporters than for hub ones given that  cubic fluctuations are normalized by the standard deviations in computing the skewness.
 
 \subsection{Correlation of local exports and in-degrees at the aggregate level}

\begin{figure}
\includegraphics[width=\columnwidth]{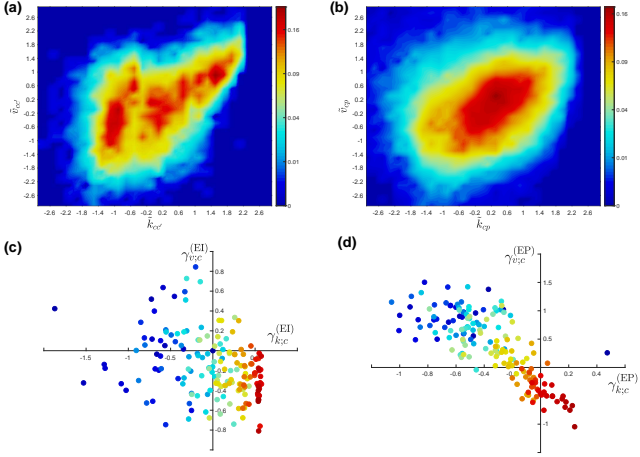}
\caption{{\bf Correlation between local in-degrees and local logarithmic exports and their skewness.}
(a) The joint probability density function (PDF) $P^{\rm (EI)} (\tilde{k},\tilde{v}) = \sum_{c,c^\prime} A_{cc^\prime} \delta(\tilde{k}_{cc^\prime}-\tilde{k})\delta(\tilde{v}_{c\nu}-\tilde{v})/\sum_{c,c^\prime} A_{cc^\prime}$ of standardized local in-degrees and local logarithmic exports in the EI network. The orientation of the red (high density) region to the upper right indicates a positive correlation. 
(b) The joint PDF  $P^{\rm (EP)} (\tilde{k},\tilde{v}) = \sum_{c,p} A_{cp} \delta(\tilde{k}_{cp}-\tilde{k})\delta(\tilde{v}_{cp}-\tilde{v})/\sum_{c,p} A_{cp}$.
(c) Scatter plot of the skewness of local logarithmic exports to different destinations $\gamma_{v;c}^{\rm (EI)}$ versus the skewness $\gamma_{k;c}^{\rm (EI)}$ of local in-degrees in the EI network. Color varies with the out-degree of $c$ as in Fig.~\ref{fig:indegreedist}. 
(d) Scatter plot of the skewness of local logarithmic exports of different products $\gamma_{v;c}^{\rm (EP)}$ versus the skewness $\gamma_{k;c}^{\rm (EP)}$ of local in-degrees in the EP network. 
} 
\label{fig:corr}
\end{figure}

At the aggregate level, the local export values of destinations/products and their in-degrees are correlated. The Pearson correlation coefficient between the aggregated standardized values $\{\tilde{v}_{c\nu}|c, \nu\in \mathtt{I}_c\}$ and $\{\tilde{k}_{c\nu}|c, \nu\in \mathtt{I}_c\}$ is 0.51 and 0.29 for EI and EP, respectively, in year 1980. Here we introduced the standardized local in-degrees $\tilde{k}$ for exporter $c$ defined by
\begin{equation}
k_\nu = \langle k\rangle_c + \sigma_{k;c} \tilde{k}_{c\nu}.
\end{equation}
 The joint distributions of the standardized local in-degrees and export in Fig.~\ref{fig:corr} (a) and (b) also support their positive correlations. 

If the local export value to/of a destination/product is strongly correlated with the in-degree of the latter, then one can expect 
\begin{equation}
\tilde{v}_{c\nu}\simeq  \tilde{k}_{c\nu}.
\label{eq:strongcorr}
\end{equation}
Then the two skewness should be identical or at least similar $\gamma_{v;c} = \langle \tilde{v}^3\rangle_c \simeq  \langle \tilde{k}^3\rangle_c =  \gamma_{k;c}$.  In Fig.~\ref{fig:corr} (c) and (d), however, one can see that they are quite different and some of them have even different signs. Moreover, $\gamma_{v;c}^{\rm (EP)}$ shows a significant negative correlation with $\gamma_{k;c}^{\rm (EP)}$. Such discrepancy between the two skewness shows that Eq.~(\ref{eq:strongcorr}) needs correction.  In the next section, we present an improved version of Eq.~(\ref{eq:strongcorr}) which relates the two skewness  correctly and leads us to understand  at a deeper level the  skewness of local logarithmic exports.

\section{Roles of in-degrees and fluctuations in local exports}
\label{sec:decomposing}

\subsection{Decomposing local exports}

\begin{figure}
\includegraphics[width=\columnwidth]{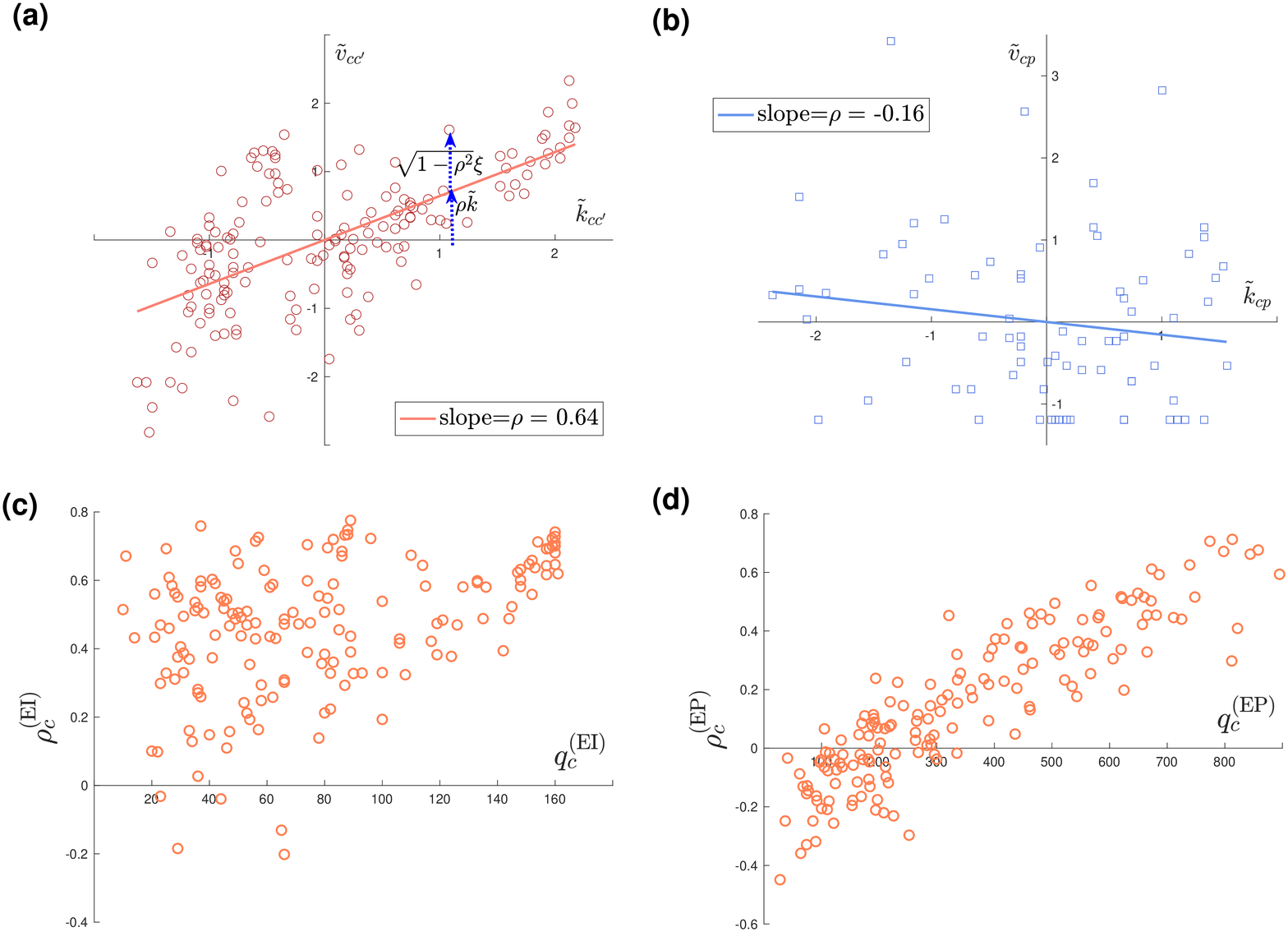} 
\caption{{\bf Formulating local exports as functions of local in-degrees for given exporter.}
(a) Standardized logarithmic exports $\{\tilde{v}_{cc^\prime}|c^\prime \in \mathtt{I}_c\}$ to different destinations $c^\prime$'s from a selected country $c$ versus their standardized in-degrees $\{\tilde{k}_{cc^\prime}|c^\prime \in \mathtt{I}_c\}$ in 1980. The slope $\rho_c$ of the fitting line is equal to their Pearson correlation coefficient. The deviation of $\tilde{v}_{cc^\prime}$ from the prediction by the in-degree, $\rho_c \tilde{k}_{cc^\prime}$, is set to $\sqrt{1-\rho^2} \xi_{cc^\prime}$, by which the random noise $\xi_{cc^\prime}$ is defined.
(b) $\{\tilde{v}_{cp}\}$ versus $\{\tilde{k}_p\}$ for a selected country $c$ in 1980. 
(c) Pearson correlation coefficients $\rho_c^{\rm (EI)}$ between the logarithmic exports to different destinations from an exporter $c$ and their in-degrees versus the out-degree of $c$. 
(d) Pearson correlation coefficients $\rho_c^{\rm (EP)}$ between the logarithmic exports of different products from an exporter $c$ and their in-degrees versus the out-degree of $c$. 
} 
\label{fig:decomposing}
\end{figure}

 The standardized local exports to/of and in-degrees of destinations/products are shown as examples for two selected countries  in Fig.~\ref{fig:decomposing} (a) and (b). The slope of the best fitting line of the data points $\{(\tilde{k}_{c\nu}, \tilde{v}_{c\nu})|\nu\in \mathtt{I}_c\}$ corresponds to their Pearson correlation coefficient, which  differs from exporter to exporter. Also we find that the data points fluctuate considerably deviating from the fitting line. Such fluctuation is different in magnitude and sign depending on the target - destination/product - for given exporter. We take the sum of all the factors contributing to local exports except for the in-degrees as a {\it random noise} and represent the local export to/of a destination/product as 
 \begin{equation}
\tilde{v}_{c\nu} = \rho_c \tilde{k}_{c\nu} + \sqrt{1-{\rho_c}^2} \xi_{c\nu},
\label{eq:decomposing}
\end{equation}
where  $\rho_c^{\rm (EI)}$ and $\rho_c^{\rm (EP)}$ are the Pearson correlation coefficients between local exports  and in-degrees given by
\begin{equation}
\rho_c= \langle \tilde{k} \tilde{v}\rangle_c = {1\over q_c} \sum_{\nu\in \mathtt{I}_c} \tilde{k}_{\nu} \tilde{v}_{c\nu}, 
\label{eq:rho}
\end{equation}
and $\xi_{cc^\prime}$ and $\xi_{cp}$ are random noises of zero mean and unit strength satisfying 
\begin{align}
\langle \xi\rangle_c &={1\over q_c} \sum_{\nu \in \mathtt{I}_c} \xi_{c\nu} = 0, \nonumber\\
\langle \tilde{k} \xi \rangle_c &={1\over q_c} \sum_{\nu \in \mathtt{I}_c} \tilde{k}_\nu \xi_{c\nu} = 0,\nonumber\\
\langle  \xi^2 \rangle_c &={1\over q_c} \sum_{\nu \in \mathtt{I}_c} \xi_{c\nu}^2 = 1.
\label{eq:xistat}
\end{align}
The relations in Eq.~(\ref{eq:xistat}) are derived by using Eq.~(\ref{eq:rho}), $\langle \tilde{k}\rangle_c = \langle \tilde{v}\rangle_c=0$, and $\langle \tilde{k}^2\rangle_c = \langle \tilde{v}^2\rangle_c = 1$. The graphical meaning and measurment of $\rho$'s and $\xi$'s are presented in Fig.~\ref{fig:decomposing} (a).   All the quantities in Eq.~(\ref{eq:decomposing}) can be measured empirically, which advances greatly our study as will be presented in the remaining part. We expect that the statistics of random noise $\xi_{c\nu}$ encodes the relevant factors mainly stemming from the exporter $c$ while the standardized in-degree $\tilde{k}$ reflect those from the destination/product. They are independent of each other in that their covariance $\langle \tilde{k}\xi\rangle_c$ vanishes, but as we will show below, they have non-zero higher-order joint cumulants.

From the dependence of $\rho_c$ on the out-degree $q_c$, more significant in the EP than the EI network [Fig.~\ref{fig:decomposing} (c) and (d)], we see that non-hub exporters determine their export values to/of destinations/products mainly by referring to their own circumstances encoded by the noise term  while  the local exports from hub exporters depend strongly on the popularity of destinations/products. 

\subsection{Joint third-order cumulants of in-degree and random noise}

\begin{figure}
\includegraphics[width=\columnwidth]{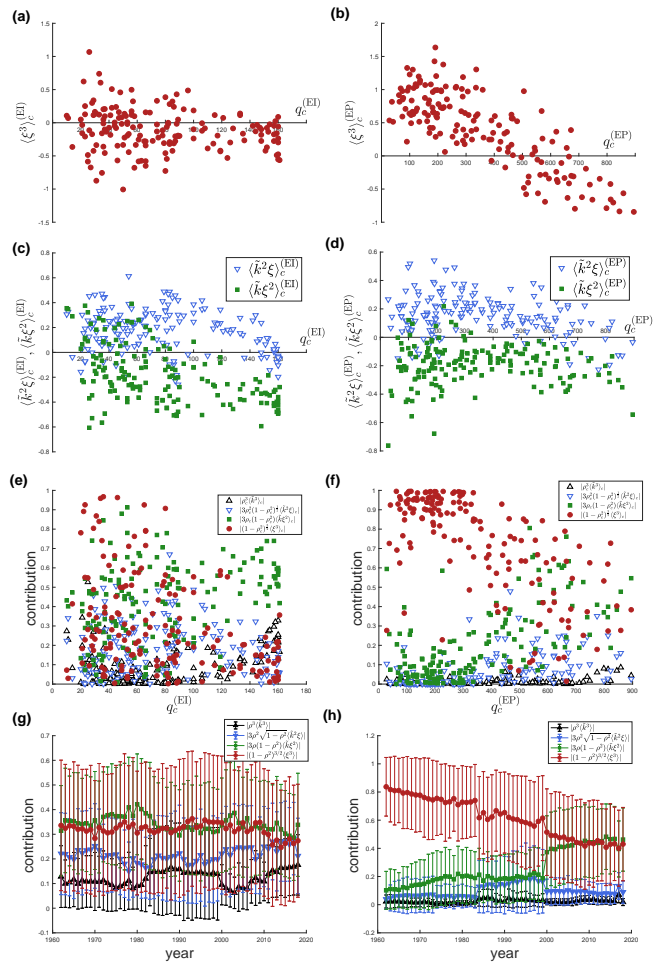} 
\caption{{\bf Third-order cumulants of local in-degrees and random noises and their relative magnitudes.} (a-f) are obtained from the data of year 1980. 
(a) The skewness of random noises $\langle \xi^3\rangle_c^{\rm (EI)}$ in the local exports from each exporter $c$ to different destinations versus the out-degree $q_c^{\rm (EI)}$.
(b) The skewness of random noises $\langle \xi^3\rangle_c^{\rm (EP)}$ in the local exports from exporter $c$ of different products  versus the out-degree $q_c^{\rm (EP)}$.
(c) Third-order joint cumulants $\langle \tilde{k}^2 \xi\rangle_c^{\rm (EI)}$ and $\langle \tilde{k} \xi^2\rangle_c^{\rm (EI)}$ versus $q_c^{\rm (EI)}$.
(d) $\langle \tilde{k}^2 \xi\rangle_c^{\rm (EP)}$ and $\langle \tilde{k} \xi^2\rangle_c^{\rm (EP)}$ versus $q_c^{\rm (EP)}$.
(e) Relative magnitudes of the four third-order cumulants of local in-degrees and random noises in the EI network.
(f) The same plot as (e) for the EP network.
(g) The averages of the relative magnitudes of the third-order cumulants  in the EI networks as functions of time. Errorbars represents the standard deviations. 
(h) The same plot as (g) for the EP networks. } 
\label{fig:contribution}
\end{figure}

Using Eq.~(\ref{eq:decomposing}), one can evaluate the skewness of local exports as
\begin{align}
\gamma_{v;c} &= \rho_c^3 \langle \tilde{k}\rangle_c + 3 {\rho_c}^2 \sqrt{1-{\rho_c}^2} \langle \tilde{k}^2 \xi\rangle_c \nonumber\\
&+ 3(1-{\rho_c}^2) \rho_c \langle \tilde{k}\xi^2 \rangle_c + (1-{\rho_c}^2)^{3/2} \langle \xi^3\rangle_c,
\label{eq:gammadecomp}
\end{align}
where four third-order cumulants, $\langle \tilde{k}^3\rangle_c, \langle \tilde{k}^2 \xi\rangle_c, \langle \tilde{k}\xi^2\rangle_c$, and $\langle \xi^3\rangle_c$ can be measured empirically and capture  asymmetry in the distribution of $\tilde{k}$'s and $\xi$'s. The four cumulants therefore represent four different modes to generate asymmetry in the local export distribution. We have seen that $\gamma_{k;c}=\langle \tilde{k}^3\rangle_c$ depends on $q_c$ in the opposite way to how $\langle \tilde{v}^3\rangle_c$ does, which we now understand can happen if $\rho_c$ is sufficiently small  as $\langle \tilde{k}^3\rangle_c$ is weighted by $\rho_c^3$ in Eq.~(\ref{eq:gammadecomp}).  $\tilde{k}$ and $\xi$ have zero covariance $\langle \tilde{k}\xi\rangle_c=0$ but we find empirically that they have non-zero third-order joint cumulants $\langle \tilde{k}^2\xi\rangle_c$ and $\langle \tilde{k}\xi^2\rangle_c$ [Fig.~\ref{fig:contribution} and \ref{fig:4termssign}].  In principle, $\langle \tilde{k}^2 \xi\rangle_c$ is positive (negative) when the data points of $(\tilde{k}, \tilde{v})$ tend to be located above (below) the fitting line $\tilde{v}^{\rm (fit)} = \rho \tilde{k}$ when $\tilde{k}^2$ is large. If  $(\tilde{k}, \tilde{v})$ tends to spread more (less) as $\tilde{k}$ increases, the joint cumulant $\langle \tilde{k}\xi^2\rangle$ will be positive (negative). 

The skewness of the random noise $\xi$ is expected to make a considerable contribution to the whole skewness of local export, for the magnitudes and the $q_c$-dependence of $\langle \xi^3\rangle_c$ are empirically close to those of $\langle \tilde{v}^3\rangle$ [Fig.~\ref{fig:contribution} (a) and (b)]. It suggests that the statistics of the random noise $\xi$, whose nature mainly depends on the exporter itself, may govern the statistics of local exports. However, the joint cumulants $\langle \tilde{k}^2\xi\rangle_c$ and $\langle \tilde{k}\xi^2\rangle_c$ are not negligible [Fig.~\ref{fig:contribution} (c) and (d)] but even make bigger contributions to the whole skewness $\langle \tilde{v}^3\rangle_c$ for some exporters $c$ [Fig.~\ref{fig:contribution} (e) and (f)]. We measure the contributions of the four third-order cumulants to the skewness of local exports by the ratio of the absolute values of the  four terms in Eq.~(\ref{eq:gammadecomp}), $|\rho^3\langle \tilde{k}^3\rangle_c|,|3 {\rho_c}^2 \sqrt{1-{\rho_c}^2} \langle \tilde{k}^2 \xi\rangle_c|,
|3(1-{\rho_c}^2) \rho_c \langle \tilde{k}\xi^2 \rangle_c|, |(1-{\rho_c}^2)^{3/2} \langle \xi^3\rangle_c|$ to their sum $|\rho_c^3 \langle \tilde{k}\rangle_c| + |3 {\rho_c}^2 \sqrt{1-{\rho_c}^2} \langle \tilde{k}^2 \xi\rangle_c| +
|3(1-{\rho_c}^2) \rho_c \langle \tilde{k}\xi^2 \rangle_c| + |(1-{\rho_c}^2)^{3/2} \langle \xi^3\rangle_c|$. Therefore the sum of those four contributions is one ($100\%$) for each exporter.  As expected, the contribution of the skewness of in-degrees is not significant, $9.0\pm 9.0\%$ to $\gamma_{v;c}^{\rm (EI)}$ and $1.0\pm 1.7\%$ to $\gamma_{v;c}^{\rm (EP)}$ in year 1980 [Fig.~\ref{fig:contribution} (e) and (f)].  On the other hand, the contribution of the skewness of the random noise $ \langle \xi^3\rangle_c$ is $31\pm 25\%$ and $73\pm 26\%$ to $\gamma_{v;c}^{\rm (EI)}$ and $\gamma_{v;c}^{\rm (EP)}$, respectively. Interestingly, the contribution of a joint third-order cumulant $\langle \tilde{k}\xi^2\rangle_c$ makes a large contribution, $42\pm 20\%$ and $20\pm 19\%$  to $\gamma_{v;c}^{\rm (EI)}$ and $\gamma_{v;c}^{\rm (EP)}$ respectively  in year 1980. Across time [Fig.~\ref{fig:contribution} (g) and (h)], $\langle \xi^3\rangle_c$ contributes about $32\%$ and $\langle \tilde{k}\xi^2\rangle$ $34\%$ to $\gamma_{v;c}^{\rm (EI)}$. For $\gamma_{v;c}^{\rm (EP)}$, time variations are stronger than for $\gamma_{v;c}^{\rm (EI)}$; $\langle \xi^3\rangle_c$ makes the leading contribution $40 \sim 84\%$ and the next-leading contribution is made by $\langle \tilde{k}\xi^2\rangle_c$, ranging $10\%$ to $48\%$. 

The empirically measured third-order cumulants allows us to better understand the skewness of local exports. $\langle \xi^3\rangle_c$ and $\langle \tilde{k}\xi^2\rangle_c$ together contribute $66\pm 4\%$  and $88\pm 5\%$ to $\gamma_{v;c}^{\rm (EI)}$ and $\gamma_{v;c}^{\rm (EP)}$, respectively. They tend to decrease with the out-degree of exporter and are thus responsible for the out-degree-dependence of $\gamma_{v;c}$. As such, it is mainly via the two cumulants $\langle \tilde{k}\xi^2\rangle_c$ and $\langle \xi^3\rangle_c$ that an exporter and its partners/products affect the skewness of their local logarithmic exports. 

Moreover, the analysis based on Eq.~(\ref{eq:gammadecomp}) elucidates the nature of the export strategies of individual countries. As shown in Fig.~\ref{fig:contribution} (c) and (d), $\langle \tilde{k}\xi^2\rangle_c$ is negative for most exporters. In 1980, 126 among 161 countries have $\langle \tilde{k}\xi^2\rangle_c^{\rm (EI)} \leq 0$ and 152 countries have $\langle \tilde{k}\xi^2\rangle_c^{\rm (EP)} \leq 0$. Such negative  $\langle \tilde{k}\xi^2\rangle_c$ is observed  for many countries throughout the considered period [Fig.~\ref{fig:4termssign}], and means that the spread of local exports is weaker for larger $\tilde{k}$. As such, a country determines its export value to a popular destination or of a popular product by more referring to its in-degree than to/of an unpopular destination/product, for the latter of which  the circumstances of the exporter itself can be important. Also we find that  $\langle \tilde{k}^2\xi\rangle_c$ is positive for many countries, which implies that the local export to/of quite a popular or quite an unpopular destination/product tends to be larger than the average. See Fig.~\ref{fig:skewnesschematic} for graphical examples. All these results illuminate how the export values of an exporter country are allocated to destinations/products, and combined with the different connection patterns between hub and non-hub exporters,  illuminate the universal and the non-universal aspects of the strategies of countries in international trade. 

\section{Summary and Discussion}
\label{sec:discussion}

We have investigated how a country divides its export value over destinations and products by analyzing the local logarithmic export distributions. They are close to the Gaussian distribution, but exhibit significant skewness. We studied the origin and the implications of such skewness for the export strategies of countries in international trade.   The local export per destination/product is not constant but large for hub exporters and small for non-hub exporters, which contributes to the variation of the skewness of local exports with exporters. We presented the formula relating the local export value to/of a destination/product to the in-degree of the latter and a random noise, which are measurable empirically  and thereby allow us to decipher the origin of the skewness of local exports and to understand quantitatively the export strategies of countries. We showed that the statistics of the random noise and its third-order correlation with in-degrees are responsible for the skewness of local exports and that the local exports to/of more popular destination/products are governed more by their in-degrees than the export to/of less popular ones are.

 Our study aimed at understanding the connection topology and the local link-weight distributions in weighted networks, here the world trade networks, and therefore our methods and results can be applied to other weighted complex networks including natural, information, social, and artificial networks.  The triangular relationships involving exporters, importers, and products can be represented by hypergraphs~\cite{PhysRevE.79.066118}, and generalizing our analysis for them is in progress.
  
\begin{acknowledgements}
We thank Su Do Yi for valuable comments.  This work was supported by the National Research Foundation of Korea (NRF) grants funded by the Korean Government (No. 2019R1A2C1003486) and a KIAS Individual Grant (No. CG079901) at Korea Institute for Advanced Study.
\end{acknowledgements}


\appendix

\renewcommand{\thesection}{A\arabic{section}}
\renewcommand{\theequation}{A\arabic{equation}}
\renewcommand{\thefigure}{A\arabic{figure}}
\renewcommand{\thetable}{A\arabic{table}}
\setcounter{figure}{0}    
\setcounter{equation}{0}

\section{Data-sets}
\label{sec:datasets}

In this study, the NBER-UN data-set in the period $1962\leq t\leq 1999$~\cite{NBERw11040} and the World Trade Organization (WTO) data-set for $2000\leq t\leq  2018$~\cite{wto2020} are compiled to obtain the set of export values $\{V_{cc'p}(t)|c,c^\prime\in \mathtt{C}(t), p\in \mathtt{P}(t)\}$ for $1962\leq t\leq 2018$.  Here $\mathtt{C}(t)$ is the set of countries exporting or importing at least one product category and $\mathtt{P}(t)$ is the set of trade product categories exported by at least one country in year $t$. To classify trade products,  the 4-digit standard international trade classification (SITC)~\cite{sitc4} is used in Ref.~\cite{NBERw11040}. The WTO data-sets~\cite{wto2020} are based on the Harmonized System (HS)~\cite{hs2017} as the product classification scheme, and we also select its 4-digit version to make the total number of product categories similar to that of the 4-digit SITC. These two schemes of product classification can be mapped to each other~\cite{unstats2020}.

We remark that the data-sets  in the period 1984 to 1999 contain some virtual product categories inserted to fill some gaps in the total values of trade across different levels of classification~\cite{NBERw11040}. The NBER-UN data-sets record trade over 1000 US dollars. Yet these differences depending on data source and time period do not affect the main results of the present study.


\begin{figure}
\includegraphics[width=\columnwidth]{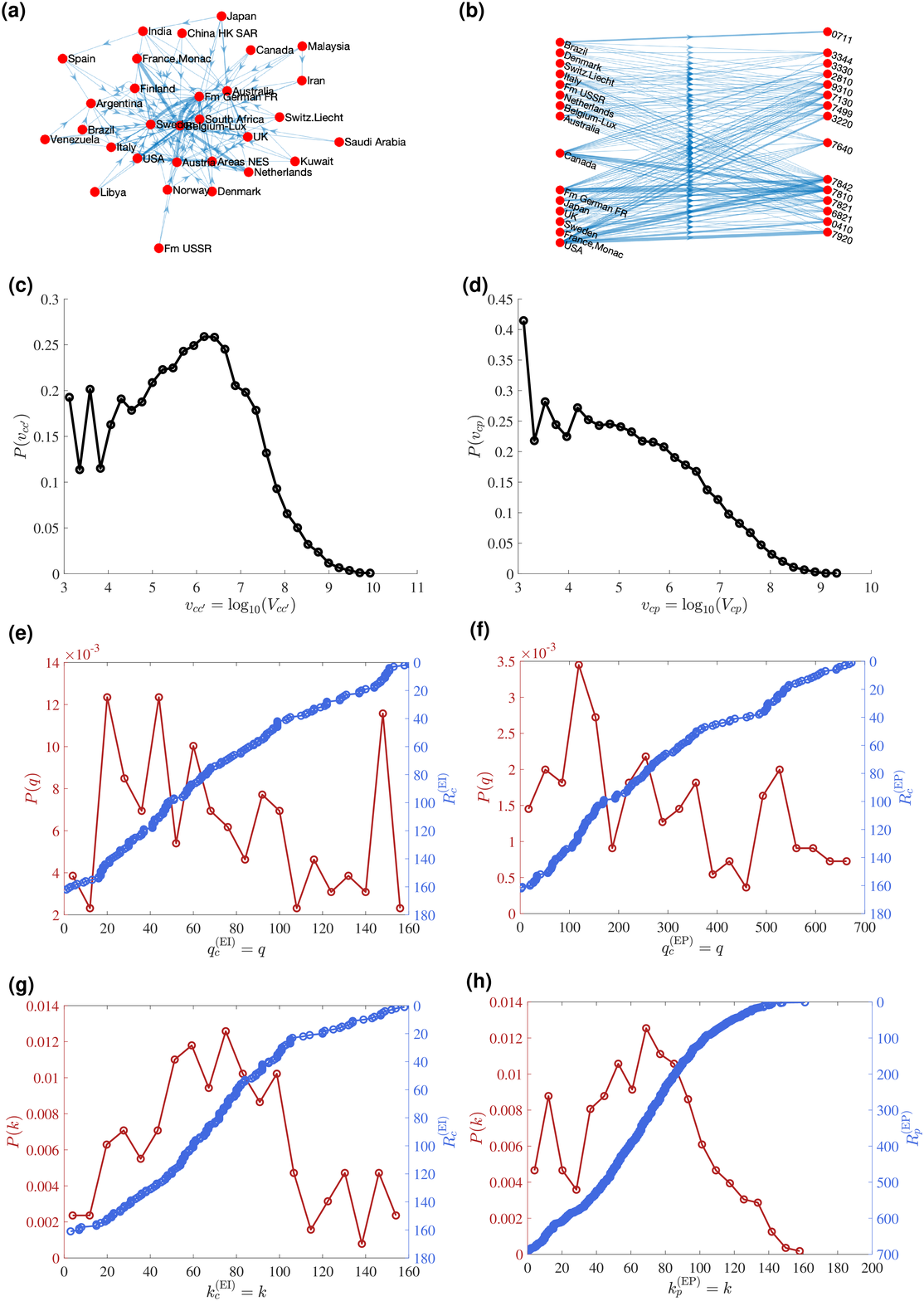}
\caption{{\bf Trade networks in 1970 and their global degree and link-weight distributions. }
(a) A subnetwork of the EI network in 1970, consisting of  30 countries having the largest (out$+$in)-degrees connected by  141 links of standardized local weight (logarithmic export value) larger than 1, i.e., $\tilde{v}> 1$. 
(b) A subnetwork of the EP network of 15 countries having the largest out-degrees and 15 products having the largest in-degrees connected by 133 links of $\tilde{v}>1$.
(c) Global distribution of the logarithm of exports  $P^{\rm (EI)} (v) = {\sum_{c,c^\prime} A_{cc^\prime} \delta(v_{cc^\prime} - v) \over \sum_{c,c^\prime} A_{cc^\prime}}$ in the EI network with $A_{cc^\prime}$ the adjacency matrix. 
(d) Global distribution of the logarithm of exports  $P^{\rm (EP)} (v)$ in the EP network.
(e) Global out-degree distribution $P^{\rm (EI)}(q) = {\sum_c \delta(q_c^{\rm (EI)} - q) \over \sum_c 1}$ and the plot of the rank $R_c^{\rm (EI)}$ versus the out-degree $q_c^{\rm (EI)}$ of each exporter $c$ in the EI network. 
(f)  Global out-degree distribution $P^{\rm (EP)}(q)$ and the plot of the rank $R_c^{\rm (EP)}$ versus the out-degree $q_c^{\rm (EP)}$ of exporter $c$ in the EP network. 
(g) Global in-degree distribution $P^{\rm (EI)}(k) = {\sum_{c} \delta(k_{c}^{\rm (EI)} - k) \over \sum_{c} 1}$ and the plot of the rank $R_c^{\rm (EI)}$ versus the in-degree $k_c^{\rm (EI)}$ of importer $c$ in the EI network. 
(f)  Global in-degree distribution $P^{\rm (EP)}(k) = {\sum_{p} \delta(k_{p}^{\rm (EP)} - k) \over \sum_{p} 1}$ and the plot of the rank $R_p^{\rm (EP)}$ versus the in-degree $k_p^{\rm (EP)}$ of product $p$ in the EP network. 
 }
\label{fig:network}
\end{figure}

\section{Significance of skewness and its correlation with out-degree}
\label{sec:sigcorr}


\begin{figure}
\includegraphics[width=\columnwidth]{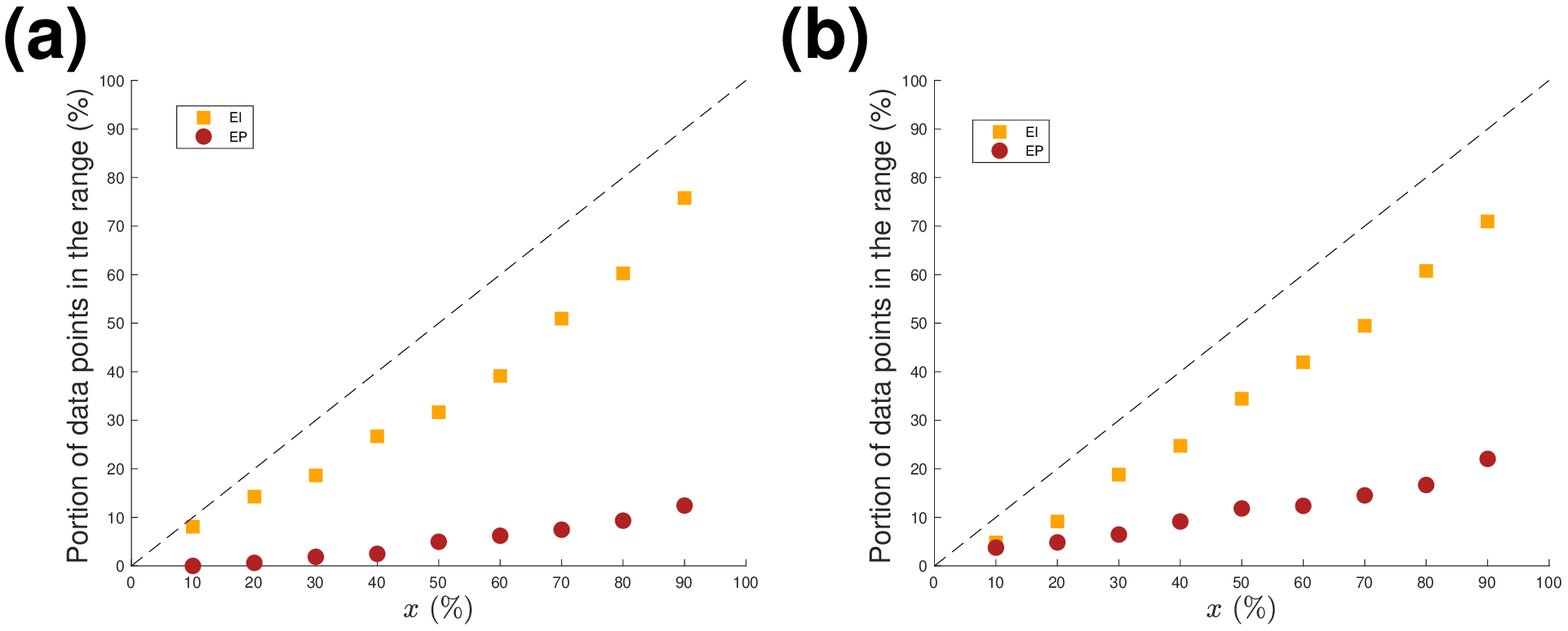} 
\caption{{\bf Significance of skewness of local logarithmic exports.}
(a) The fraction of exporters whose true skewness remain in the middle $x(\%)$ range of the random-version skewness in 1980. Squares and circles are for the EI and the EP network.
(b) The same plot as (a) in year 2010. 
} 
\label{fig:sig}
\end{figure}

To check the significance of the skewness of local logarithmic exports, we compare the true skewness with the values of the skewness from  $\mathcal{N}=1000$ realizations of weight-shuffled EI and EP networks. The out-degrees of each exporter $q_c^{\rm (EI)}$ and $q_c^{\rm (EP)}$ do not change by this randomization. Then, for each exporter $c$ we are given $\mathcal{N}$ values of the random-version skewness $\{\gamma_{v;c}^{{\rm (EI; random)} (\ell)}|\ell=1,2,\ldots, \mathcal{N}\}$ in the EI network and $\{\gamma_{v;c}^{{\rm (EP; random)} (\ell)}|\ell=1,2,\ldots, \mathcal{N}\}$ in the EP network, and we check  whether the true skewness $\gamma_{v;c}^{\rm (EI)}$ and $\gamma_{v;c}^{\rm (EP)}$ belong to the middle $x (\%)$ range of those random-version skewness values. Repeating this  for all exporters, we obtain the fraction $y(\%)$ of exporters whose true skewness are included in  their respective  middle $x(\%)$ ranges of the random-version skewness. If $y$ as a function of $x$ remains close to $x$, then one can say that true skewness is not distinguished from the random-version ones and therefore the true skewness is not a significant one. On the other hand, if $y$ is far smaller than $x$, then it means that the true skewness tends to be outside the range expected in the randomized networks, demonstrating the significance of the observed skewness in a sense that it is hardly expected in randomized networks.

We found that the latter is the case. In Fig.~\ref{fig:sig}, we present the plot of $y$ versus $x$ for the EI and the EP network in year 1980 and 2010. $y$ is far below the line $y=x$ in case of the EP network. Also for the EI network, $y$ is  smaller than $x$ though the difference is not so impressive as for the EP network. These results suggest that the empirically-observed skewness of local logarithmic exports are significant.


\begin{figure}
\includegraphics[width=\columnwidth]{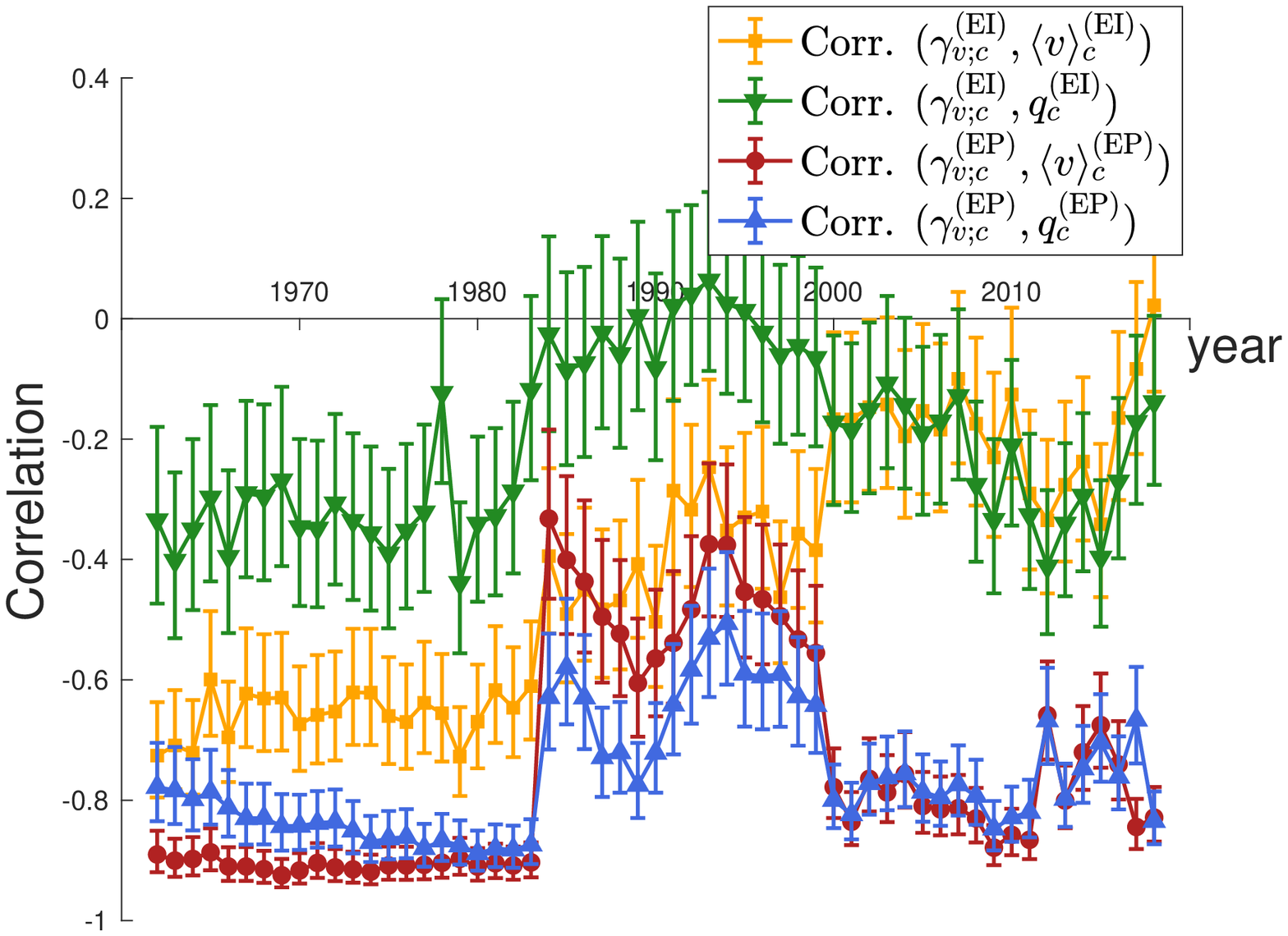} 
\caption{{\bf Time-dependent correlation of the skewness with the average of local logarithmic exports and with the out-degree of exporter.} The Pearson correlation coefficient between the skewness of local logarithmic exports $\gamma_{v;c}^{\rm (EI,EP)}$ and the out-degree $q_c^{\rm (EI,EP)}$ of exporter or the average local logarithmic export $\langle v\rangle_c^{\rm (EI,EP)}$ as a function of time $t$. 
} 
\label{fig:corrskewdiversity}
\end{figure}

\begin{figure}
\includegraphics[width=\columnwidth]{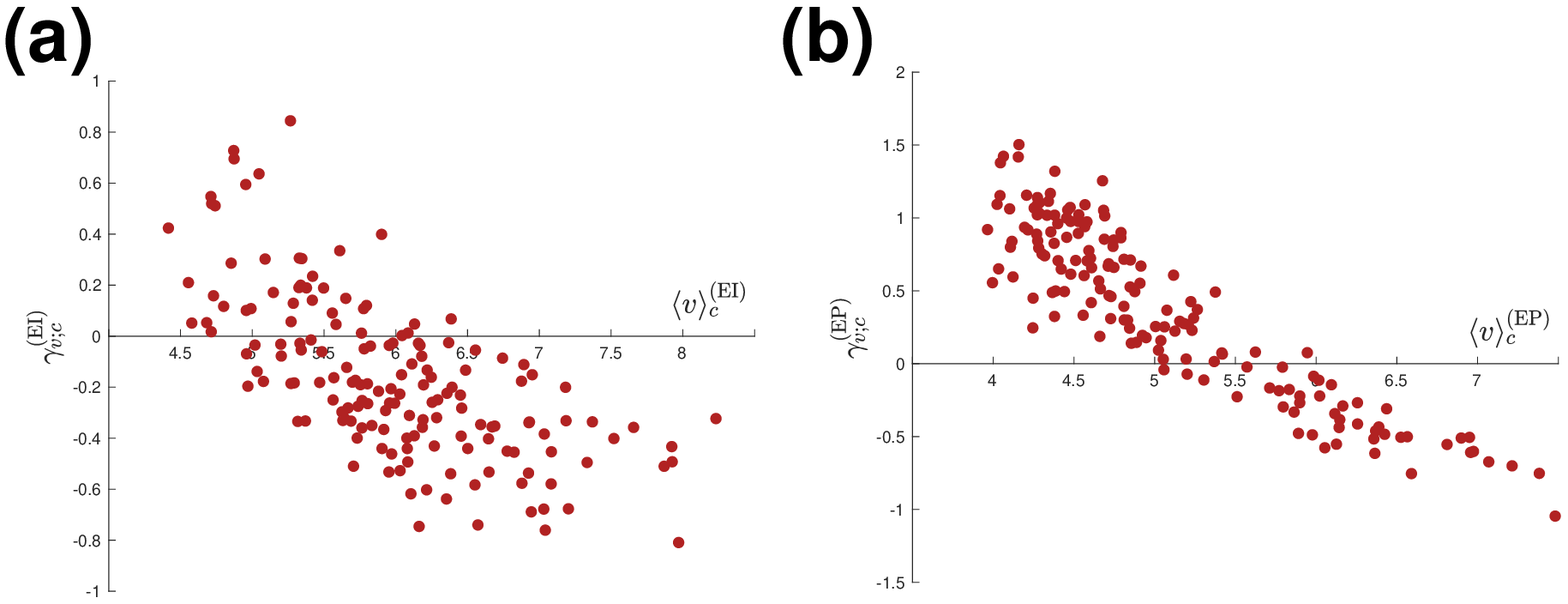} 
\caption{{\bf Plot of the skewness versus the average of local logarithmic exports.} (a) Plot of the skewness of local logarithmic exports $\gamma_{v;c}^{\rm (EI)}$ versus the average of local logarithmic exports $\langle v\rangle_c^{\rm (EI)}$ in the EI network in year 1980. (b)  The same plot as (a) in the EP network. }
\label{fig:skewness_mvc}
\end{figure}

\begin{figure}
\includegraphics[width=\columnwidth]{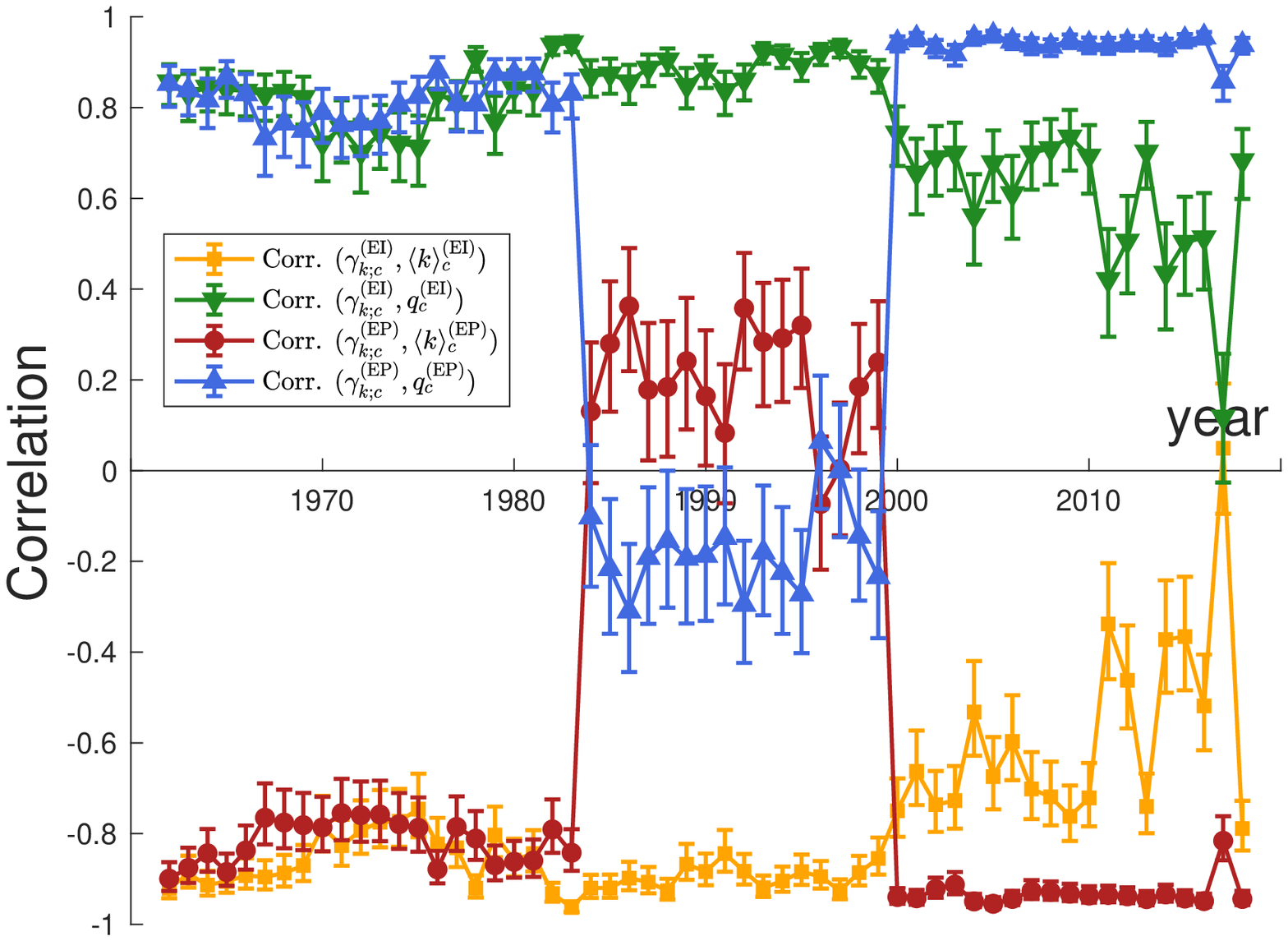} 
\caption{{\bf Correlation of skewness of local in-degrees with the out-degree of exporter and the  average   local in-degree.} The Pearson correlation coefficient between the skewness of local in-degrees $\gamma_{k;c}^{\rm (EI,EP)}$ and the out-degree $q_c^{\rm (EI,EP)}$ of exporter or the average local in-degree $\langle k\rangle_c^{\rm (EI,EP)}$ as a function of time $t$. Correlations are  observed robustly in the studied period, except for the correlations of $\gamma_{k;c}^{\rm (EP)}$ with $q_c^{\rm (EP)}$ and with $\langle k\rangle_c^{\rm (EP)}$ which change sign  in the period 1984-1999 presumably owing to the virtual product categories added artificially [See Appendix~\ref{sec:datasets}].}
\label{fig:corrskewindeg}
\end{figure}

\begin{figure}
\includegraphics[width=\columnwidth]{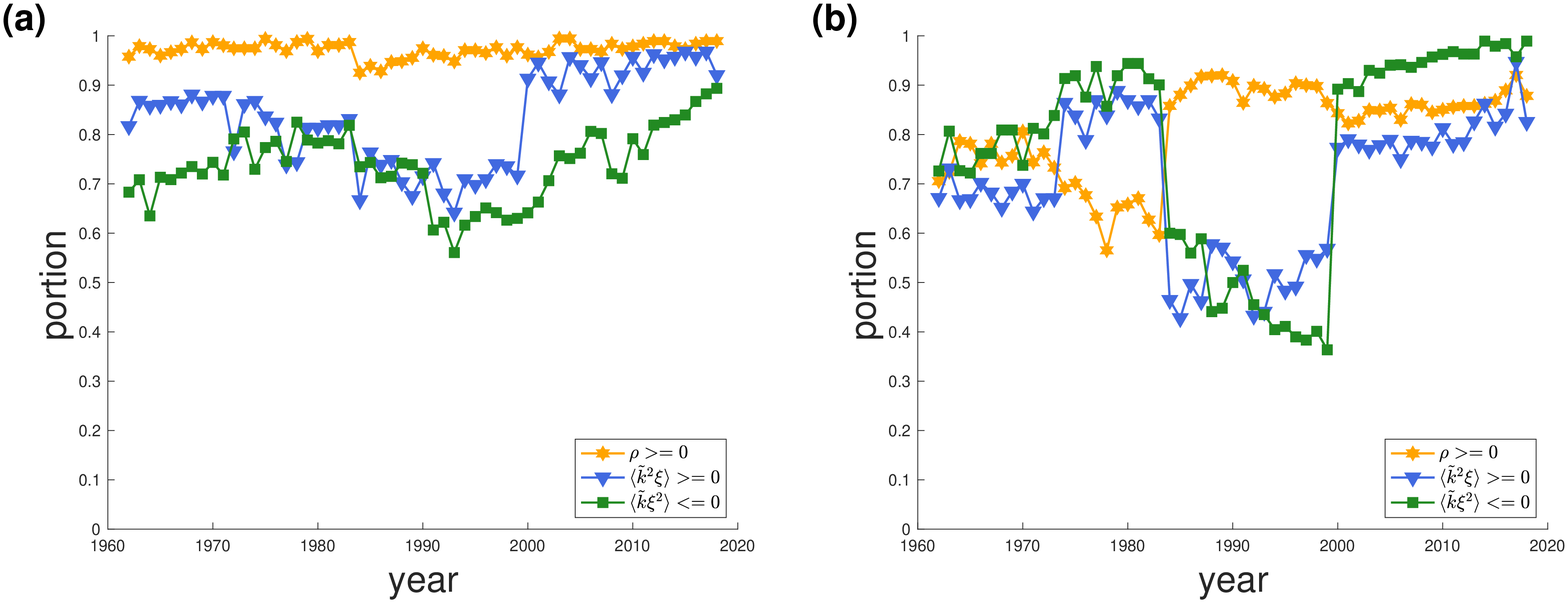} 
\caption{{\bf Robustness against time of the sign of the correlation coefficient and third-order cumulants. }
(a) The portions of exporters showing a non-negative correlation coefficient between local logarithmic exports and local in-degrees ($\rho_c^{\rm (EI)}\geq 0$), a non-negative cumulant $\langle \tilde{k}^2\xi\rangle_c^{\rm (EI)}\geq 0$, and another non-positive cumulant $\langle \tilde{k}\xi^2\rangle_c^{\rm (EI)}\neq 0$ of local in-degrees and random noises are shown respectively for the EI network as functions of time.
(b) The same plots as (a) for the EP network. 
} 
\label{fig:4termssign}
\end{figure}

\begin{figure}
\includegraphics[width=\columnwidth]{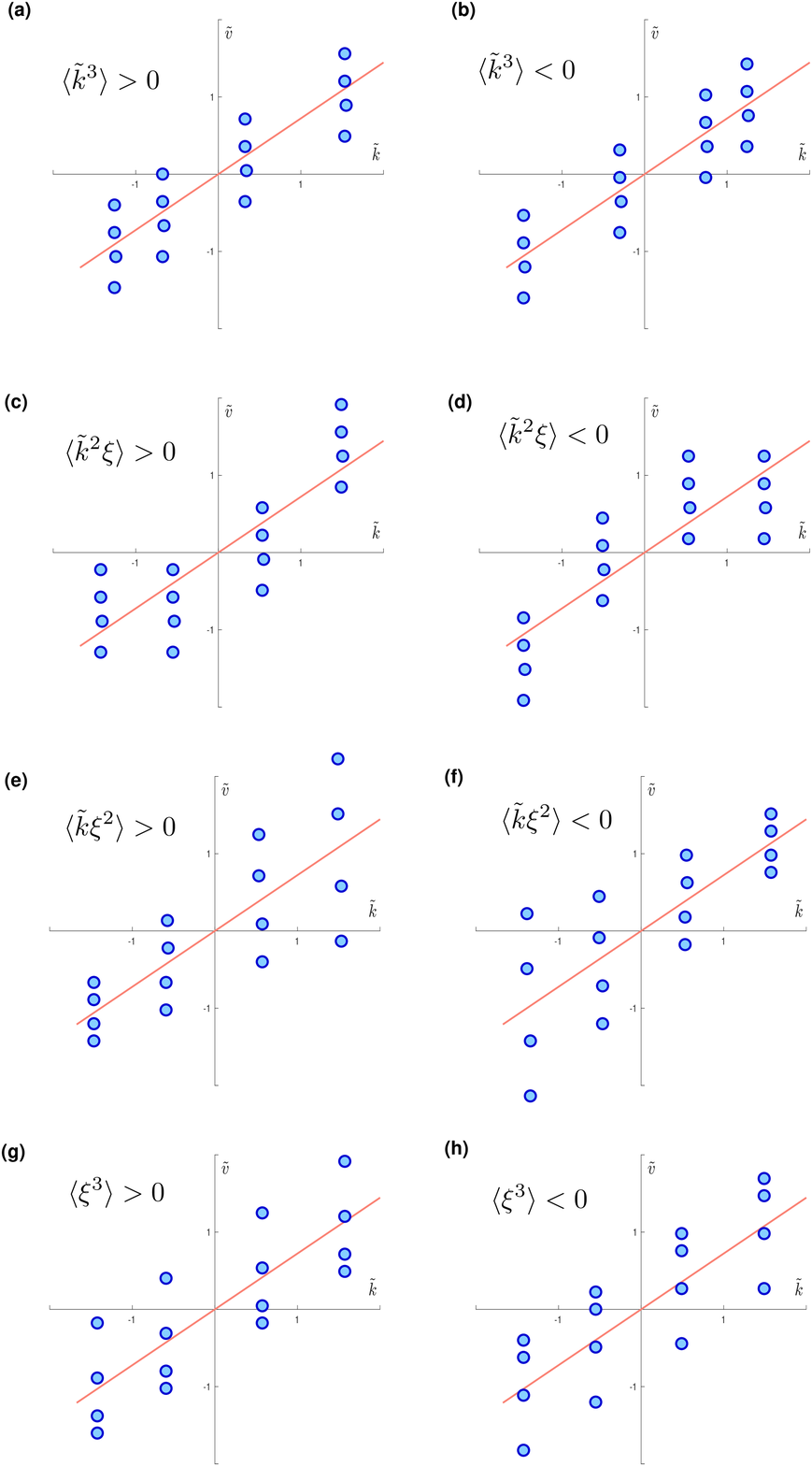} 
\caption{{\bf Schematic distribution of $\{(\tilde{k}, \tilde{v})\}$ corresponding to positive or negative third-order cumulants.} Each panel shows an example of the distribution of the data points of standardized local exports and in-degrees  $\{(\tilde{k}, \tilde{v})\}$ exhibiting a given positive or negative cumulant. The solid line represents $\tilde{v}=\rho \tilde{k}$ with $\rho=1$ for simplicity. Therefore the deviation of the data points from the line represents the random nooise $\xi$. Note that $\langle \tilde{k}\rangle=\langle \xi \rangle=0$ and $\langle \tilde{k}^2\rangle = \langle \xi^2\rangle=1$. 
} 
\label{fig:skewnesschematic}
\end{figure}


%

\end{document}